\documentclass[british]{rsproca_arxiv}

\usepackage[T1]{fontenc}
\usepackage[utf8]{inputenc}
\usepackage{amstext}
\usepackage{amssymb}
\usepackage{graphicx}
\usepackage{esint}

\linkfalse
\webfigfalse
\makeatletter
\usepackage{epstopdf}

\makeatother

\usepackage{babel}
\begin{document}

\title{Generation of surface waves by an underwater moving bottom:\\
 Experiments and application to tsunami modelling}

\author{Timothée Jamin$^{1}$, Leonardo Gordillo$^{1}$, Gerardo Ruiz-Chavarría$^{2}$,
Michael Berhanu$^{1}$, Eric Falcon$^{1}$}

\address{$^{1}$Univ Paris Diderot, Sorbonne Paris Cité, MSC, UMR 7057 CNRS,
F-75013 Paris, France, \\
$^{2}$Facultad de Ciencias, Universidad Nacional Autónoma de México,
04510 México D. F., México}

\corres{Leonardo Gordillo\\
\email{leonardo.gordillo@univ-paris-diderot.fr}}

\keywords{tsunami generation, gravity surface waves}

\subject{fluid mechanics, wave motion, geophysics}
\begin{abstract}
We report laboratory experiments on surface waves generated in a uniform
fluid layer whose bottom undergoes a sudden upward motion. Simultaneous
measurements of the free-surface deformation and the fluid velocity
field are focused on the role of the bottom kinematics in wave generation.
We observe that the fluid layer transfers bottom motion to the free
surface as a temporal high-pass filter coupled with a spatial low-pass
filter. Both filter effects are usually neglected in tsunami warning
systems. Our results display good agreement with a prevailing linear
theory without fitting parameter. Based on our experimental data,
we provide a new theoretical approach for the rapid kinematics limit
that is applicable even for non-flat bottoms: a key step since most
approaches assume a uniform depth. This approach can be easily appended
to tsunami simulations under arbitrary topography.
\end{abstract}
\maketitle

\begin{fmtext}

\end{fmtext}

\section{Introduction}

Most tsunamis are triggered by sudden displacements of the seabed
during earthquakes. To predict tsunami hazards in real time, actual
warning models require, first and foremost, data of the free-surface
waveform in the open seas. Buoy networks dedicated to detect tsunamis
may provide direct measurements of wave heights at fixed positions
across the oceans \cite{2012PApGe.tmp...42M}. When buoy records are
unavailable, a faster indirect method based on fault and seismic data
is used by default. The seabed displacement is computed numerically
from the fault slip using Okada's model \cite{Okada:1985uy}, and
then transferred to the ocean free surface. Unfortunately, this technique
often underestimates the surface wave amplitude (\emph{e.g. }for the
2004 Sumatra-Andaman tsunami \cite{2006SciAm.294a..56G}).

Several reasons have been proposed to explain this bias \cite{2006EP&S...58..185G,Maeda:2011cp,Tanioka:2001vu,Dutykh:2010hv,Synolakis:2002kp,Okal:2004ju,2008JFM...598..107S,Geist:2007ds},
including the seabed-kinematics role during an earthquake (\emph{i.e.}
its spatiotemporal features) \cite{Geist:2007ds,1997Sci...278..598S,Geist:1998vc,Todorovska:wf}.
Bottom displacement is considered to be instantaneous if its typical
rise time is small compared to the time scale of the generated waves
at the free surface \cite{Geist:1998vc}. Most earthquakes meet this
condition, although other remarkable tsunamigenic events barely satisfy
it: for instance, in two of the hugest tsunami ever registered, the
bottom displacements were noticeably slow \cite{1972PEPI....6..346K}.
Unfortunately, many numerical codes used in warning systems neglect
seabed kinematics (\emph{e.g.} the MOST \cite{Titov:1997vx} and the
TUNAMI \cite{UNESCO:1997vt}). Instead, they use a transfer model
that simply translates the source bottom final deformation to the
ocean surface. Although available, numerical simulations that suitably
do consider bed-sea kinematic coupling during displacements have high
computation costs (cf. \cite{Ichiye:1958ug,Aida:1969tq,Kervella:2007dc,Kakinuma:2009tu})
and require to know the bottom kinematics a priori. Strikingly, even
if the deformation happens instantaneously, the free-surface displacement
is not equal to the bottom one \cite{Kajiura:1963tn,Dutykh:2006jh}.

Laboratory experiments dealing with the influence of bed-uplift kinematics
in tsunami generation are rare and have been based on measurements
of the free-surface deformation \cite{Takahasi:1962tt,1973JFM....60..769H},
providing limited information about the fluid dynamics. Velocity measurements
in the bulk are even rarer and only concern landslide-triggered tsunamis
\cite{Fritz:2003du,Fritz:2003iu}. Furthermore, many laboratory experiments
have been performed in channels and thus tend to overlook the three-dimensional
(3D) geometry of real scenarios \cite{1973JFM....60..769H,Fritz:2003du,Fritz:2003iu,2014EL....10534004V}.

In this Letter, we analyse experimentally and theoretically the hydrodynamic
coupling between the bottom and the free-surface motion in a 3D fluid
layer, focusing on the role that the bottom kinematics plays in wave
generation. For this purpose, we performed combined measurements of
the free-surface deformation and the fluid velocity field. Our results
are then compared with a linear theory \cite{Hammack:1972wf}. We
also provide a generalized framework that can be applied systematically
to enhance tsunami warning-systems simulations.

\section{Experimental Setup}

\begin{figure}
\begin{centering}
\includegraphics[width=12cm]{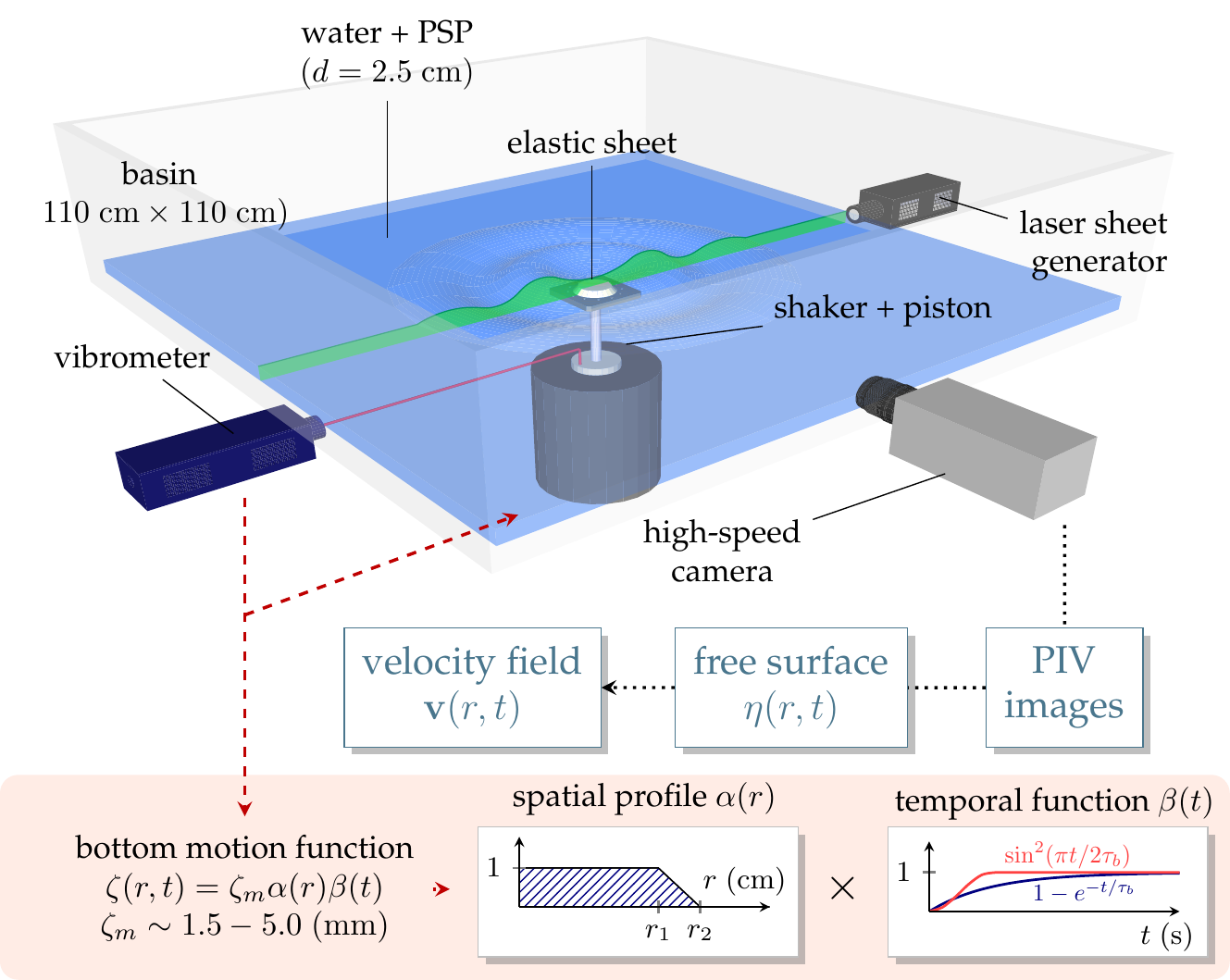} 

\par\end{centering}

\caption{Main: Experimental setup. A layer of water is contained in a basin
where a shaker and a piston vertically deform an elastic sheet placed
at the bottom centre. The piston motion is recorded using a laser
vibrometer. Images from a sectional cut of the fluid are obtained
using a laser sheet generator and a high-speed camera. The bottom
dimensionless spatial profile $\alpha\left(r\right)$ and the time
displacement function $\beta\left(t\right)$ are also displayed.\label{fig:setup}}
\end{figure}

We performed our experiments in a $110\times110\times30\,\mathrm{cm^{3}}$
Plexiglas basin filled with water to a depth of $h=2.5\,\mathrm{cm}$.
A circular region (radius $r_{2}=3.25\,\mathrm{cm}$) was carved in
the bottom centre and covered with a stretched elastic sheet. The
sheet is deformed by means of a solid flat circular piston ($r_{1}=2.5\,\mathrm{cm}$)
placed beneath the membrane and attached to an electromechanical shaker
(see Fig.~\ref{fig:setup}). As a result of the setup geometry, the
bottom vertical motion can be described as a separable spatiotemporal
function with circular symmetry, $\zeta\left(r,t\right)=\zeta_{m}\alpha\left(r\right)\beta\left(t\right)$,
where $\zeta_{m}$ is the maximal bottom deformation; $\alpha\left(r\right)$
is the spatial profile along the radial horizontal coordinate $r$
{[}see Fig.~\ref{fig:setup} (inset){]} and $\beta\left(t\right)$
is the displacement time function. The latter was arbitrary chosen
to be an exponential rise, $\beta_{\exp}\left(t\right)=1-e^{-t/\tau_{b}}$,
or a half-sine one, $\beta_{\sin}\left(t\right)=\sin^{2}\left[\pi t/\left(2\tau_{b}\right)\right]$
if $t\leq\tau_{b}$ or 1 if $t>\tau_{b}$; where $\tau_{b}$ is defined
as the rise time. To achieve this, the shaker input signal was determined
by exploiting the bottom velocity records from a laser Doppler vibrometer.
Our system can be used to study rise times from 10 to $500\,\mathrm{ms}$,
and upward bottom amplitudes from $1.5$ to $5\,\mathrm{mm}$. Typical
bottom velocities vary from $1$ to $30\,\mathrm{cm\cdot s^{-1}}$.
The basin extent  was chosen to avoid wave reflections on the lateral
walls during the generation process. 

The velocity field in the bulk during bottom and surface deformations
is obtained using Particle Image Velocimetry (PIV). A laser sheet
passing through the basin centre illuminated a vertical slice of water
seeded with $50$-$\mathrm{\mu m}$ polyamide particles (see Fig.~\ref{fig:setup}).
This provided an imaging region of $71\times30\,\mathrm{mm^{2}}$
($1600\times692\,\mathrm{pixels}$) that was recorded at $500\,\mbox{Hz}$
during $1\,\mathrm{s}$. Since the system is axisymmetric, these measurements
build a 3D picture of the flow. The surface of the water layer was
blown with more particles to create an identifiable line on the images
for detection. The free-surface vertical deformation $\eta\left(r,t\right)$
was then obtained by applying a Radon transform algorithm on the images
\cite{2011ExFl...51..871S}. Finally, we applied a PIV grid-refining
scheme \cite{1997ExFl...23...20W} using an average correlation method
\cite{Meinhart:2000wc} on ten experimental runs for each set of parameters.
All the data used throughout this article is available at a public
repository at \cite{repository-tsunami}.

\begin{figure}
\begin{centering}
\includegraphics{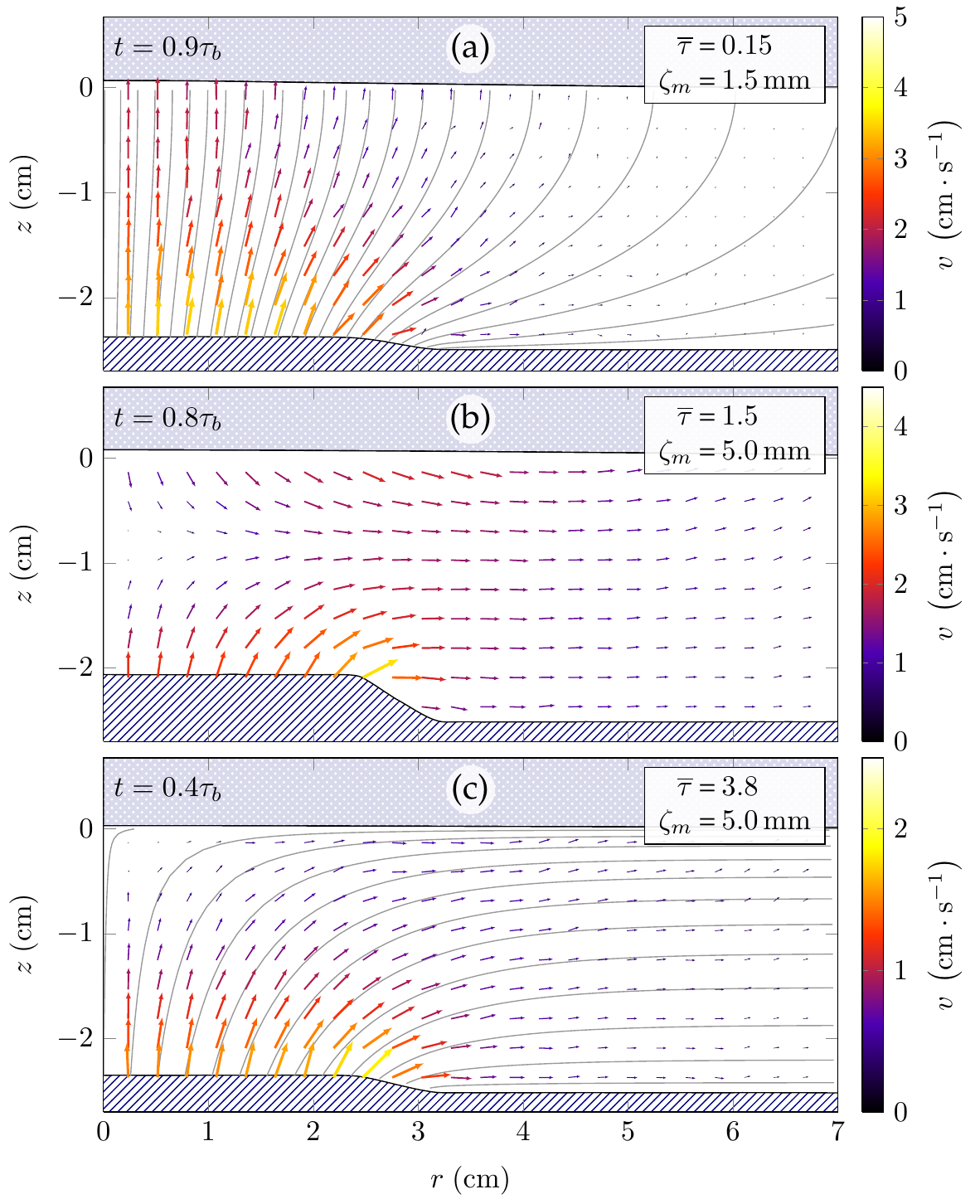} 
\par\end{centering}

\caption{Experimental velocity fields in the water during half-sine bottom
displacements for three typical $\overline{\tau}$ values. The streamlines
(set of grey curves) in (a) and (c) were computed numerically using
an asymptotic model for the $\overline{\tau}\ll1$ and $\overline{\tau}\gg1$
cases (see explanation in text). In all cases, the free-surface deformation
is significantly smoother than the bottom one.\label{fig:field}}
\end{figure}

The two time scales in our experiment are the bottom rise time $\tau_{b}$
and the typical time of the generated waves $\tau_{w}$. We defined
$\tau_{w}$ as the semi-period of the wave,\emph{ i.e.} the time between
the first maximum and minimum of the water surface deformation at
the basin centre ($r=0$). In our experiment, $\tau_{w}\simeq130\,\mathrm{ms}$
is the same for any displacement time function $\beta$ and rise time
$\tau_{b}$ (see results below). This value is related to the dominant
wavelength of the generated wave $\lambda_{w}\approx10\,\mathrm{cm}$
(according to acquired images), through the dispersion relation $\tau_{w}=\pi/\sqrt{gk_{w}\tanh k_{w}h}$.
Accordingly, the experimental time ratio, defined as $\overline{\tau}=\tau_{b}/\tau_{w}$
varies between 0.08 and 4. The relevance of the time ratio in tsunami
generation was noticed by Hammack \cite{1973JFM....60..769H}, who
suitably identified three wave-response regimes to bottom deformations:
impulsive ($\overline{\tau}\ll1$), transitional ($\overline{\tau}\sim1$),
and creeping ones ($\overline{\tau}\gg1$).

\section{Results and discussion}

Within this classification, we display in Fig.~\ref{fig:field} three
characteristic snapshots of the generation velocity fields for half-sine
type displacements. The vertical coordinate is denoted as $z$ such
that at rest, the free surface matches $z=0$, and the bottom, $z=-h$.
When $\overline{\tau}\ll1$, we observe an upward global motion during
the bottom uplift. Indeed, the velocity field just below the free
surface is vertical {[}see Fig.~\ref{fig:field}(a){]} as predicted
in \cite{2000PhFl...12.2819T}. Gravity-wave propagation  starts remarkably
after the end of the bottom motion as shown in videos %
\footnote{See Supplemental Material at {[}URL{]} for velocity-field videos of
the runs depicted in Fig.~\ref{fig:field}.%
}. When $\overline{\tau}\sim1$, the flow resembles that of Fig.~\ref{fig:field}(a)
at short times. However, before the bottom motion ends, waves start
to propagate radially from the generation region: an oscillating flow
occurs right beneath the free surface {[}see Fig.~\ref{fig:field}(b){]}.
In this case, both bottom deformation and wave propagation occur simultaneously
suggesting that the bottom kinematics affects induced waves. For $\overline{\tau}\gg1$,
the free surface remains mostly stationary and accordingly, the vertical
component of the velocity vanishes when approaching the free surface
{[}see Fig.~\ref{fig:field}(c){]}. In this stage, the outward flow
reminds that of a moving bottom in presence of a fixed boundary at
$z=0$.

\begin{figure}
\begin{centering}
\includegraphics{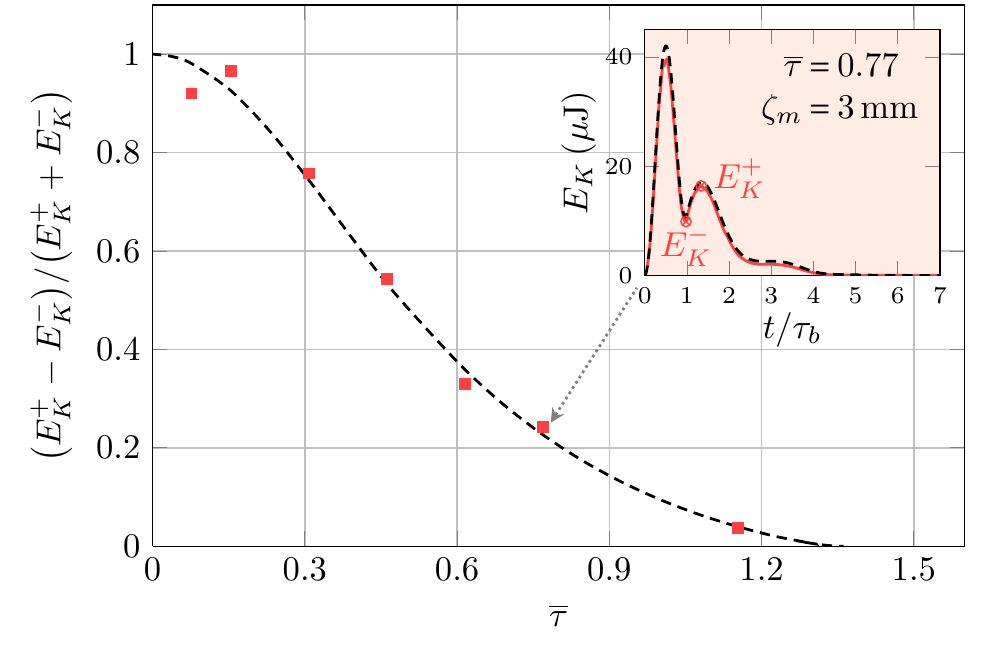} 
\par\end{centering}

\caption{Inset: Experimental (solid line) and theoretical (dashed line) kinetic
energy $E_{K}$ versus time. $E_{K}^{-}$ is the local minimum of
kinetic energy near $t=\tau_{b}$ and $E_{K}^{+}$ is the local maximum
of kinetic energy due to the wave propagation. Main: Experimental
(squares) and theoretical (dashed line) contrast of kinetic energy
versus $\overline{\tau}$.\label{fig:deltaE}}
\end{figure}

To quantify the transition between the slow and rapid regimes, we
compute the kinetic energy from the fluid velocity field. Figure~\ref{fig:field}
shows that the region $r<7\,\mathrm{cm}$ contains most of the kinetic
energy during the bottom deformation. As shown in Fig.~\ref{fig:deltaE}
(inset), the kinetic energy within this volume, $E_{K}$, captures
also the main temporal features of the motion (see also \cite{Dutykh:2009fa}).
The bottom uplift induces an intense first maximum of $E_{K}$. As
the bottom stops afterwards, a local minimum $E_{K}^{-}$ appears
and later, a second maximum $E_{K}^{+}$ emerges induced by wave propagation.
We define the contrast of kinetic energy as $\left(E_{K}^{+}-E_{K}^{-}\right)/\left(E_{K}^{+}+E_{K}^{-}\right)$.
As shown in Fig.~\ref{fig:deltaE}, the contrast is close to unity
for $\overline{\tau}\ll1$: the liquid can be considered as motionless
at the end of the bottom deformation $\left(E_{K}^{-}\approx0\right)$,
with its velocity being negligible compared to those due to wave propagation.
Inertia seems to be absent since no flow outlasts the bottom motion:
the liquid layer and the bottom behaves like a single block. For larger
$\overline{\tau}$, the wave propagation begins while the bottom is
still moving so the energy contrast decreases to zero. Furthermore,
for $\overline{\tau}\gtrapprox1.4$, the contrast is not any more
defined because the bottom deformation and the wave propagation overlap
so much that $E_{K}^{-}$ and $E_{K}^{+}$ do not exist at all. This
shows that the energy contrast depicts well the transition between
rapid and slow scenarios.

\begin{figure}
\begin{centering}
\includegraphics{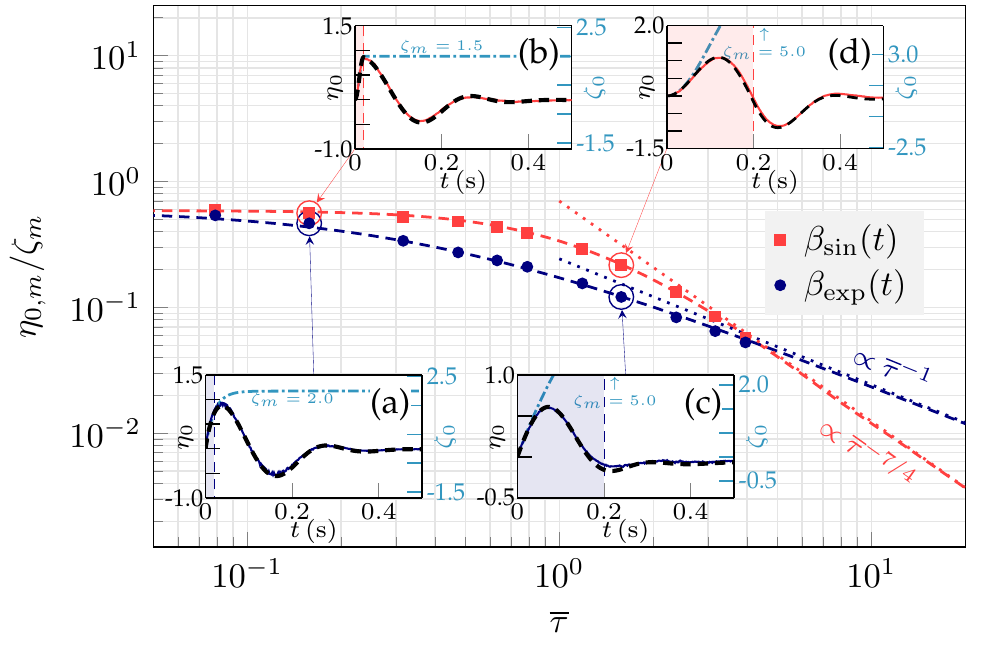} 
\par\end{centering}

\caption{Insets: Bottom and free-surface deformations in $\mathrm{mm}$ at
$r=0$, $\zeta_{0}$ and $\eta_{0}$, versus time ($\zeta_{0}$ in
dash-dotted lines; $\eta_{0}$ experiments in solid, theory in dashed
lines) for exponential (a and c) and half-sine (b and d) bottom displacements.
The vertical dashed lines represent $t=\tau_{b}$ for each run. The
wave time scale $\tau_{w}$ is found to be $130\,\mathrm{ms}$ in
all the cases. Main: Dimensionless free-surface maximal elevation
at $r=0$, against $\overline{\tau}$ for different bottom displacements
(see legend). Symbols are experimental data, dashed lines are theoretical
results and dotted lines are asymptotic behaviours $\left(\overline{\tau}\gg1\right)$.
\label{fig:eta_t}}
\end{figure}

Since the fluid velocity field is coupled with the free-surface deformation,
both quantities share similar spatiotemporal features. In Fig.~\ref{fig:eta_t},
we depict the bottom and the free-surface elevations at $r=0$, $\zeta_{0}$
and $\eta_{0}$, as a function of time. We observe in all cases that
the free surface and the bottom are synchronized at the beginning
of the motion. For $\overline{\tau}\ll1$ this is true throughout
the bottom uplift and regardless of the displacement time function
$\beta\left(t\right)$ as time satisfies $t<\tau_{b}\ll\tau_{w}$
{[}see Figs.~\ref{fig:eta_t}(a)-\ref{fig:eta_t}(b){]}. Besides,
the subsequent stage is independent of the displacement-time history
of the bottom. Contrariwise, for $\overline{\tau}\gtrsim\text{1}$,
exponential and half-sine bed displacements induce free-surface responses
that not only differ from rapid ones, but also from one another {[}see
Figs.~\ref{fig:eta_t}(c)-\ref{fig:eta_t}(d){]}, \emph{e.g.} the
negative part of $\eta_{0}$ is more pronounced for the half-sine
case. This evidences that for $\overline{\tau}\gtrsim\text{1}$ the
generated-wave shape  depends on the nature of $\beta\left(t\right)$
as well as on its typical time $\tau_{b}$, which confirms that the
bottom kinematics is crucial in non-impulsive wave generation.

To understand more precisely its role, we plot in Fig.~\ref{fig:eta_t}~(main)
the dimensionless maximal elevation of the free surface at $r=0$,
$\eta_{0,m}/\zeta_{m}$, as a function of the time ratio $\overline{\tau}$.
As expected, $\eta_{0,m}$ decreases with $\overline{\tau}$ and converges
to the same asymptote for $\overline{\tau}\ll1$ independently of
the nature of $\beta\left(t\right)$. For $\overline{\tau}\gg1$,
we observe two different behaviours: $\eta_{0,m}$ decreases as $\tau_{b}^{-1}$
for exponential bottom displacements and as $\tau_{b}^{-7/4}$ for
half-sine ones. This differs from 1D experiments where a $\tau_{b}^{-1}$
power law fits both cases \cite{1973JFM....60..769H}. To summarize,
when motion is transferred from the bottom to the free surface, the
fluid layer behaves as a temporal high-pass filter (cut-off at $\overline{\tau}^{-1}\approx1$).

The experimental data displayed in Figs.~\ref{fig:deltaE}-\ref{fig:eta_t}
as well as the spatial profiles (not shown herein), are all found
to be in good agreement with theoretical curves without fitting parameters.
These were calculated using the axisymmetric version of Hammack's
tsunami-generation theory \cite{Hammack:1972wf} Capillary and viscosity
effects can be neglected since $\lambda_{w}\gg2\pi\sqrt{\gamma/\left(\rho g\right)}$
($\gamma$ is the surface tension and $\rho$ the fluid density) and
$r_{1}\gg\sqrt{\nu\tau_{b}}$ ($\nu$ is the kinematic viscosity).
The flow is assumed to be irrotational and incompressible ($c_{s}\tau_{b}\gg\left\{ \zeta_{m},h,r_{1}\right\} $
where $c_{s}$ is the sound speed in water \cite{1999PCEB...24..437N}).
Hence the system can be expressed in terms of a velocity potential
$\phi$ that satisfies 
\begin{equation}
\nabla^{2}\phi=0\label{eq:laplace}
\end{equation}
in the bulk. The experimental amplitude parameter $\zeta_{m}/h$ is
small enough to linearize boundary conditions \cite{Hammack:1972wf}.
Thus, if the bottom is initially flat, the dynamic condition at the
free surface as well as the kinematic boundary conditions can be written
as 
\begin{eqnarray}
\left.\partial_{t}\phi\right|_{z=0}+g\eta & = & 0.\label{eq:DC_top}\\
\left.\partial_{z}\phi\right|_{z=-h}-\partial_{t}\zeta & = & 0,\label{eq:KC_bottom}\\
\left.\partial_{z}\phi\right|_{z=0}-\partial_{t}\eta & = & 0.\label{eq:KC_top}
\end{eqnarray}
To solve this system of equations, we apply the Laplace transform
in $t$ to the displacement time function, $\tilde{\beta}\left(s\right)\equiv\mathcal{L}\left\lbrace \beta(t)\right\rbrace \left(s\right)$,
and the Hankel transform of zeroth order in $r$ to the spatial profile,
$\hat{\alpha}\left(k\right)\equiv\mathcal{H}_{0}\left\{ \alpha(r)\right\} \left(k\right)\equiv\int_{0}^{+\infty}\rho J_{0}\left(kr\right)\alpha\left(r\right)\,\mathrm{d}r$,
where $J_{0}$ is the zeroth order Bessel function of the first kind.
The latter is equivalent to a two-dimensional (2D) Fourier transform
under circular symmetry. Accordingly, the Hankel transform of the
free-surface deformation may be written as \cite{Hammack:1972wf}
\begin{equation}
\hat{\eta}\left(k,t\right)=\frac{\zeta_{m}\hat{\alpha}\left(k\right)}{\cosh kh}\cdot\mathcal{L}^{-1}\left\lbrace \frac{s^{2}\tilde{\beta}\left(s\right)}{s^{2}+\omega\left(k\right){}^{2}}\right\rbrace \left(k,t\right).\label{eq:eta}
\end{equation}
where $\omega(k)=\sqrt{gk\tanh kh}$ is the gravity-wave dispersion
relation. The direct and inverse Laplace transforms in Eq.~(\ref{eq:eta})
can be evaluated in closed form for both $\beta_{\exp}\left(t\right)$
and $\beta_{\sin}\left(t\right)$. Besides, the spatial transform
$\hat{\alpha}\left(k\right)$ may be computed numerically. The spatiotemporal
free-surface deformation $\eta(r,t)=\mathcal{H}_{0}^{-1}\left\{ \hat{\eta}\left(k,t\right)\right\} $
can be found likewise using a Fourier-Bessel series representation
of $\mathcal{H}_{0}^{-1}$ \cite{Arfken2005Mathematical}. The velocity
field can also be obtained by calculating the velocity potential $\phi$
through analogous formulas.

Remarkably, the first factor in Eq.~(\ref{eq:eta}) is the Hankel
transform of the final bottom deformation but modulated with a low-pass
filter, $(\cosh kh)^{-1}$, that smooths the free surface (see Fig.~\ref{fig:field}).
The second factor is spatiotemporal and relates the time $t$ (corresponding
to $s$ in the Laplace domain) with the two characteristic times:
the wave semi-period $\tau_{w}$ (corresponding to $\omega$) and
the bottom rise time $\tau_{b}$ (contained in $\tilde{\beta}\left(s\right)$).
When $t\ll\tau_{w}$, $s^{2}+\omega^{2}\sim s^{2}$, the second factor
of Eq.~(\ref{eq:eta}) becomes simply $\beta\left(t\right)$, gravity
effects vanish yielding interface elevations instantaneously equal
to the bottom low-pass-filtered deformations. This is consistent with
the behaviour observed at short times in Fig.~\ref{fig:eta_t} (insets),
where the free surface moves synchronously with the bottom. When $t\gtrsim\tau_{b}$,
$\beta(t)$ can be considered as a Heaviside function if $\overline{\tau}\ll1$.
Hence, $\tilde{\mathrm{H}}\left(s\right)=s^{-1}$, and the second
factor in Eq.~(\ref{eq:eta}) becomes a propagation term $\cos\left[\omega\left(k\right)t\right]$.
As stated by Kajiura \cite{Kajiura:1963tn}, this is equivalent to
a Cauchy-Poisson wave problem in which only the final bottom deformation
is low-pass filtered and translated to the surface as an initial condition.
Likewise, we have shown that the fluid is motionless when the bottom
motion ends. No trace from the initial motion is left. This explains
the memory loss of the bottom-displacement history observed in our
experiments.

The impulsive limit ($\overline{\tau}\ll1$) has a striking feature:
gravity plays no role during the bottom motion ($t<\tau_{b}$). Accordingly,
we can drop the gravity term in Eq.~(\ref{eq:DC_top}), so $\left.\phi\right|_{z=0}=0$
and the free surface $\eta$ decouples from Eqs.~(\ref{eq:laplace}-\ref{eq:KC_bottom}).
This yields a decoupled boundary value problem (DBVP) for the velocity
potential $\phi$. In Fig.~\ref{fig:field}(a), we depict the streamlines
obtained from solving numerically the DBVP. This numerical method
differs from the Green function approach developed in \cite{2000PhFl...12.2819T}.
The streamlines are steady for $t<\tau_{b}$. Likewise, another DBVP
can be found for the $\overline{\tau}\gg1$ case: $g\rightarrow\infty$,
hence $\left.\partial_{z}\phi\right|_{z=0}=0$, which yields the streamlines
of Fig.~\ref{fig:field}(c). For both limits, $\overline{\tau}\gg1$
and $\overline{\tau}\ll1$, the computed streamlines fit very well
the experimental velocity field {[}see Figs.~\ref{fig:field}(a)
and \ref{fig:field}(c){]}. While for $\overline{\tau}\gg1$, $\eta\simeq0$,
for initially flat bottoms undergoing impulsive uplifts ($\overline{\tau}\ll1$),
$\eta$ can be obtained from Eq.~(\ref{eq:KC_top}). This leads to
the spatial low-pass filtered results found previously. The DBVP approach
has a great advantage: it can be adapted to arbitrary-bottom-shaped
basins by simply writing the bottom condition as $\left.\partial_{z}\phi\right|_{z=-h(x,y)}=\partial_{t}\zeta$.
We strongly recommend this method as a computationally affordable
routine in actual simulations for incorporating terrain conditions
during tsunami generation.To compare our results~$\left(^{\dagger}\right)$
with real tsunami scenarios~$\left(^{*}\right)$, consider two dimensionless
parameters: the time ratio $\overline{\tau}$ and the size scale $r_{1}/h$.
For tsunamis $\left(\zeta_{m}^{*},h^{*},r_{1}^{*}\right)\sim\left(10\,\mathrm{m},5\,\mathrm{km},80\,\mathrm{km}\right)$
and $\tau_{b}^{*}\in\left[1,100\right]\,\mathrm{s}$ \cite{1973JFM....60..769H,Geist:1998vc}.
Notice that for tsunamis and our experiments, $\zeta_{m}/h\ll1$,
so the linear theory is valid, $\eta\propto\zeta_{m}$ and $\zeta_{m}$
can be rescaled out from Eq.~(\ref{eq:eta}). Concerning $\overline{\tau}$,
the tsunami range $\overline{\tau}^{*}\in\left[0.003,0.3\right]$
is located on the left-hand side of Fig.~\ref{fig:eta_t} since here
$\overline{\tau}^{\dagger}\in\left[0.08,4\right]$. The temporal high-pass
filter becomes significant for slowest scenarios ($\eta_{0,m}^{*}/\zeta_{m}^{*}$-corrections
from $10\%$ to $40\%$ for $\overline{\tau}^{*}\sim0.3$). Although
fastest tsunamis are beyond our experimental range, the asymptote
for $\overline{\tau}\ll1$ is largely attained within it. Concerning
the other parameter, we fixed $r_{1}^{\dagger}/h^{\dagger}=1$ to
highlight the spatial low-pass filtering. For accepted tsunami values
($r_{1}^{*}/h^{*}\sim16$) these filtering effects are expected to
be weak. However, recent and more direct evidence shows that tsunami
initial waveforms have a complex spatial distribution with significantly
smaller length scales: $r_{1}^{*}/h^{*}\lesssim5$ \cite{2008EP&S...60..993F,2011EP&S...63..815F}.
For a spherical-cap deformation, this yields low-pass filtering $\eta_{0,m}^{*}/\zeta_{m}^{*}$-corrections
of $10\%$ \cite{Kajiura:1963tn}. Besides, ocean depth near subduction
tsunamigenic regions varies abruptly, \emph{e.g. }$80\%$ along $50\,\mbox{km}$
in the fault crosswise direction. Such terrain geometry provides further
significant corrections \cite{2000PhFl...12.2819T}. It is at this
point where a general-bathymetry DBVP approach will become useful.

\section{Conclusions}

In conclusion, we have investigated the generation of surface waves
by an underwater moving bottom. The experiments, which included simultaneous
measurements of fluid velocity field and free-surface displacement
in an initial flat bottom configuration, display excellent agreement
without fitting parameter with a linear theory of gravity waves. Essentially,
the fluid layer transfers motion from the bottom to the free surface
as a temporal high-pass filter coupled with a spatial low-pass filter.
Transfer models that perform a simple translation as those used by
tsunami warning systems, overlook both filters effects. Supported
on measured velocity fields, we have developed an alternative guideline
for taking into account spatial filtering in impulsive bottom uplifts.
Furthermore, we can use our model to include in situ bathymetry data
at low computational cost: a key for improving tsunami simulations
in real scenarios.

\section*{Acknowledgement}

We thank A. Lantheaume and the LIED for their technical help. T. J.
was supported by the DGA-CNRS Ph.D program and L. G., by a 2012 Postdoctoral
Fellowship of the AXA Research Fund. G. R. was supported by the program
Research in Paris 2011 of the City of Paris. This research was financed
by the ANR Turbulon 12-BS04-0005.

\end{document}